# On machine learning analysis of atomic force microscopy images for image classification, sample surface recognition

I. Sokolov[1,2,3] *

[1.] Department of Mechanical Engineering, Tufts University, Medford, MA 02155, USA

[2] Department of Biomedical Engineering, Tufts University, Medford, MA 02155, USA

[3] Department of Physics, Tufts University, Medford, MA 02155, USA

* Email: Igor.Sokolov@Tufts.edu

**Atomic force microscopy (AFM or SPM) imaging is one of the best matches with machine learning (ML) analysis among microscopy techniques. The digital format of AFM images allows for direct utilization in ML algorithms without the need for additional processing. Additionally, AFM enables the simultaneous imaging of distributions of over a dozen different physicochemical properties of sample surfaces, a process known as multidimensional imaging. While this wealth of information can be challenging to analyze using traditional methods, ML provides a seamless approach to this task. However, the relatively slow speed of AFM imaging poses a challenge in applying deep learning methods broadly used in image recognition. This Prospective is focused on ML recognition/classification when using a relatively small number of AFM images, small database. We discuss ML methods other than popular deep-learning neural networks. The described approach has already been successfully used to analyze and classify the surfaces of biological cells. It can be applied to recognize medical images, specific material processing, in forensic studies, even to identify the authenticity of arts. A general template for ML analysis specific to AFM is suggested, with a specific example of the identification of cell phenotype. Special attention is given to the analysis of the statistical significance of the obtained results, an important feature that is often overlooked in papers dealing with machine learning. A simple method for finding statistical significance is also described.**

Artificial Intelligence (AI) and its central part, Machine Learning (ML), are proliferating in all areas of research and development. ML has already been introduced to microscopy. [1-3] It has been demonstrated for the electron, [4,5] Raman, [6,7] optical, [8,9] X-ray microscopies. [2,10] The importance of image recognition is hard to overestimate. It can be used to make medical diagnoses, to predict effective treatments (prognostics), identify the origin of artifacts, etc. The use of ML analysis of AFM images has already been demonstrated to detect cancer [11,12], to identify different cell phenotypes at the level of single cells. [13] The AFM technique is fundamentally different from the other imaging methods. [14-18] AFM allows not only imaging a sample surface but also obtaining a large number of physical and chemical parameters of the sample surface. [19-21] AFM allows for attaining a very high spatial resolution. Its high resolution comes with the price; AFM is extremely sensitive to multiple noises. This may result in hard-to-identify artifacts. [22] These and other factors are paramount when combining AFM with ML.

ML analysis has also been used to control image acquisition in AFM,[17,23] to improve the accuracy of nanomechanical measurements, [24,25] to help do the reconstruction of sample properties and structures that are difficult to find using classical mathematical methods, [25-29] and to control imaging. [23,30-32] These applications are typically specific to particular models and samples. For example, the AFM control depends on the intermolecular and interatomic forces acting between the AFM probe and sample. The variety of these forces, their combinations, and the environments in which the sample can be immersed is too high to make a universal feedback control based on ML. However, ML can simplify and even fully automate the AFM operation for a particular type of sample. [33] There were a few attempts to use multi-channel information that can be recorded at each pixel to recognize particular material, [34-36] and even cells [37] at each pixel. This is a classical problem of spectroscopy with the ML analysis used in multiple spectroscopies.

Image classification is a different area of application of ML in AFM. It is less instrument-dependent than the control and image improvements. [16] Compared to well-developed ML methods used for image recognition, AFM images frequently do not have readily identifiable features when imaging complex surfaces (for example, biological cells and surfaces after specific treatment or processing). Generally, image analysis can be either the recognition of specific parts within the image (not based on the pixel spectroscopy mentioned above) or the recognition of the sample based on the entire image. These are similar problems because one can always consider a zoomed part of the original image and reduce the image recognition to the zoomed area, thereby reducing the recognition of specific image components to the problem of recognition of the entire image. [38] Generally, this problem belongs to weakly supervised problems in ML, [39] which requires different methods of image analysis.

These days, image recognition is typically associated with convolution neural networks (CNN). The major problem of CNN and other deep learning neural networks is the need for a large database. AFM, being a relatively slow technique, may have difficulty generating a sufficient number of images. It is hard to build CNN for a relatively small database. It should be noted, however, that the database can be generated for training purposes using a chosen physical model. [40] If the researchers are dealing with unknown features of the sample surface responsible for classification, or there is no definite physical model available to generate a database for training, it is extremely challenging to use deep learning methods like CNN. To the best of our knowledge, there were only a few attempts to apply CNN to classify AFM images of biological cells. [16, 41]

Here, we describe ML approaches that can be applied to *a relatively small database.* These are based on non-deep-learning ML methods, such as decision trees, regression methods, and non-deep learning neural networks. This Perspective describes the important steps to combine AFM with ML for image analysis, with a particular emphasis on image classification and analysis of the robustness of the obtained results. Robustness, or statistical significance, is important but not well-developed for the ML methods. Here, we describe a rather simple method to evaluate the statistical significance of the ML classification.

**_What is AFM?_** AFM (sometimes also called scanning probe microscopy or SPM) is the youngest novel paradigm in microscopy. It was invented in 1986 [42] as a significant sibling of scanning tunneling microscopy, which was awarded the Nobel Prize in 1984. AFM can easily be understood when described as a microscopic "finger" that can gently touch a sample surface. The images of the surface are formed similar to a human feeling the surface topography by touching. This finger is called the AFM probe. Assisted by a computer, the AFM probe scans over the surface, creating an image. This analogy is quite deep. AFM can image everything that a small learning finger can "feel." For example, it can detect different materials by feeling different adhesive, rigidity, magnetic, and electric properties. It can even be used to scratch materials to draw specific images and circuits. [43] Modern AFM allows for simultaneous imaging of distributions of more than a dozen different physical parameters at scales not accessible by any other techniques. [20, 44, 45]

**_Which AFM images/channels are best suited for ML analysis?_** A wealth of different types of AFM

images (also called channels) may be confusing when deciding on which channels to use. In contrast with other microscopies, most AFM images show intensity in absolute units. For example, the classical height image shows the map of the heights of the sample surface. After subtracting a fitted plane (or just a constant), the height distribution can be treated as absolute, which can be used directly for importing in ML algorithms. Other examples of absolute values of physical properties include adhesion, deformation, energy losses due to the probe disconnection from the sample surface, various electrical and piezoelectrical properties, etc. [14, 20, 44]. These are the best types of images to use in ML.

Several other popular imaging channels show the sample properties that depend on a particular instrument and/or are strongly dependent on the imaging control parameters (feedback gains, scanning speed, load force, etc.). When working with ML methods, it is particularly important to use data collected from multiple instruments/laboratories, which can be expanded later by other researchers. Therefore, the use of such images in ML analysis would be difficult (though not impossible) due to low quantitative repeatability and/or strong dependence on a particular hardware/software implementation. As an example of such channels, the phase images recorded in the tapping (AC) mode are susceptible to many parameters of imaging and, therefore, prone to low repeatability [46, 47]. Another example is the various error images. These are the error signals of the AFM feedback, which depends on the specific hardware and software of a particular AFM. It may even strongly depend on a particular AFM scanner (if different scanners are used in the same microscope).

It is worth noting that AFM is prone to various artifacts specific to this technique. AFM images (or maps of physical properties) are formed through the force of interaction between the AFM probe and sample ("felt" by the microscopic finger). The sensitivity of this technique is so high that it can be influenced by forces comparable to the break of one hydrogen bond, which is one of the weakest intermolecular forces. Since the inter -atomic and -molecular forces are still not well understood at the nanoscale, it is not always easy to recognize possible artifacts. Furthermore, the surface properties imaged with AFM may depend on the geometry of the AFM probe. If an extremely sharp probe is required, it is impractical to guarantee the same geometry on all probes. Therefore, it is important to verify sufficient repeatability of the surface properties imaged with different AFM probes and under a range of imaging conditions.

### *AFM Image Analysis with ML*

***Similarity with other image analyses.***

The general principles of AFM image recognition with ML are undoubtedly similar to other microscopies. However, there are important differences. Let us start with the similarity. First, it is the range of questions one can answer using ML. A somewhat simplified diagram of such questions is presented in **Figure 1**. One can deal with *supervised* and *unsupervised* data analyses. "Supervised" simply means that one knows what is imaged. For example, we can study two different types of biological cells, cancerous and normal. If we know which type of cells we imaged, it is ready for the supervised ML analysis. Different types of samples are usually called classes in ML. In this example, there are two classes: cancerous cells and normal cells. The identification of these two classes is called the classification problem. The same problem could be reformulated in another way: one wants to know the probability that a cell belongs to a particular class. This type of question belongs to the regression problem. The classification algorithm should give you a binary answer (yes/no), whereas the regression algorithm gives you a range of numbers. For example, regression algorithms can be used to predict complex sample properties using indirect measurements [18]. Examples of specific ML algorithms are listed in Figure 1.

"Unsupervised" learning means that one does not have complete knowledge of what was imaged. One could try to see a possible clustering of the images with respect to particular measures (called the sample features). The sample could be known, but one cannot say for sure what exactly we will see in the image. In this case, one could answer a question about the existence of clusters within each image [34]. For example, the sample may be a known composite material that contains several different components. However, the spatial distribution of these components is unknown. Unsupervised ML can be used in this case to try to identify the presence of these five components in the collected images. (Note that a similar problem can also be successfully solved using the supervised ML [29].)

In general, the major difference between supervised and unsupervised methods can be formulated as follows. In supervised learning, one should expect to predict/identify a class. Unsupervised learning aims to get insights into new data.

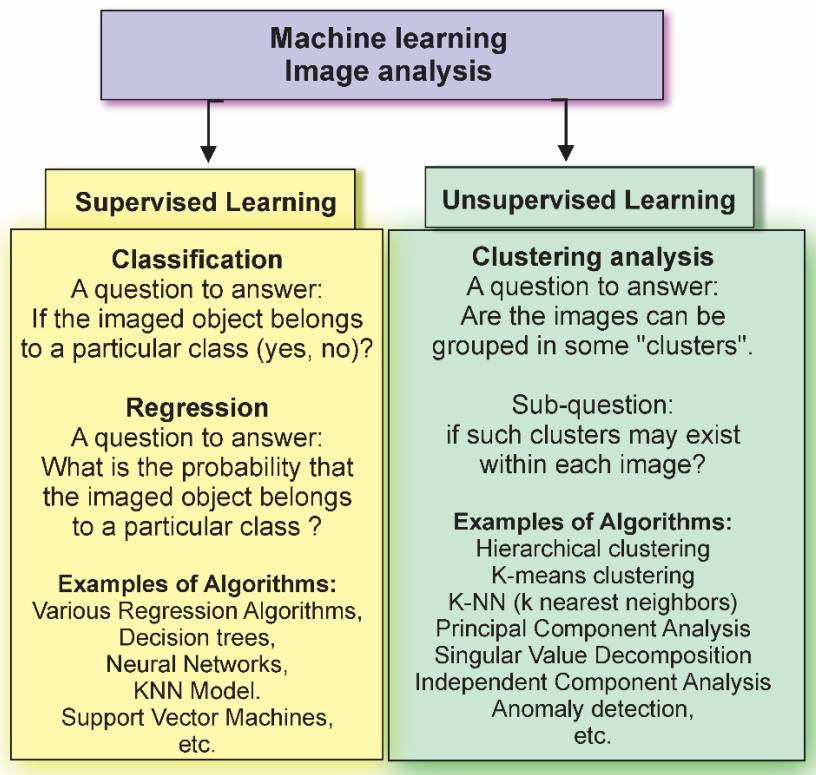

**Figure 1**. Typical problems that can be addressed using ML image analysis and examples of specific ML algorithms.

***Difference from other microscopies.***

As already mentioned, a positive, distinctive feature of the many AFM images is the absolute values of imaged sample properties. It seriously simplifies the ML analysis. There is no need for preprocessing of the data. The other distinctive feature is the imaging of multiple sample properties simultaneously. A typical ML analysis of AFM images can include the use of multiple imaging channels of the same sample surface, see the example of cell classification below.

On the negative side, AFM can be relatively slow. Although there are high-speed AFM modes, there is a concern about repeatability; the use of these images for ML has not been demonstrated. Furthermore, the number of different imaging channels in this modality is rather limited. Therefore, we will not consider this mode at the moment. Due to its slow speed, the number of images available for ML analysis can be much lower compared to other microscopies. If we follow "the rule of 10", [48] the number of measurements has to be at least ten times bigger than the number of "features" (values that characterize each image). If one takes each pixel of the image as a feature, the number of required images will be impractical (it is interesting to note an alternative approach in which the training set was generated through pure modeling [15]). It implies a mandatory step of data reduction (reduction of the dimension of the data space), which is optional in many ML approaches. This will be described in the example below.

*Recommended steps for the ML analysis of AFM images.*

Figure 2 shows a schematic of the recommended steps specific to the ML analysis of AFM images. Step 1 was briefly described above. One has to identify the imaging channels and feedback control parameters to ensure that the images are repeatable. The preferred imaging channels should record the physical parameters of the sample surface, which are the least model- and instrument-dependent. By now, these parameters include the height, adhesion, and deformation channels (recorded in a sub-resonance mode like PeakForce tapping), as well as the majority of Ringing mode channels [20]. The mechanical properties can also be included (can be imaged in AFM-nanoDMA and FT-NanoDMA modes) [44]. The choice of the feedback control parameters can be verified to obtain repeatable images by simply repeating imaging of the same surface, see, for example, ref. [49].

The need for Step 2 was described above. There are many different ways to do the data reduction. Probably the most commonly used method is convolution [16]. It is broadly used with neural networks for so-called deep learning. The description of all these methods is beyond the scope of this Communication. A specific example of this step will be given later in the text. Step 3 is rather standard for the supervised ML analysis. In the most generic way, it is presented as a set of ROC curves and the confusion matrix. In the case of unsupervised ML, there is no universal assessment of the result. Therefore, we will formulate the "quality control" of the obtained results only for supervised learning. It is presented as Step 4.

Note that despite a large number of works dealing with the ML analysis, Step 4, quality control is rarely present. We consider this as a serious drawback because ML can be a sort of black box (even if the ML algorithm is crystal clear). Frequently, the results can hardly be interpreted in terms of usual scientific logic. Therefore, this step is an important verification of the absence of possible artifacts. It can include the estimation of possible overtraining (the Achilles heel of the ML algorithms), and the statistical significance of the obtained classification. While a rigorous and comprehensive description of Step 4 is beyond the scope of this work, we give a rather detailed example of Step 4 in the next section.

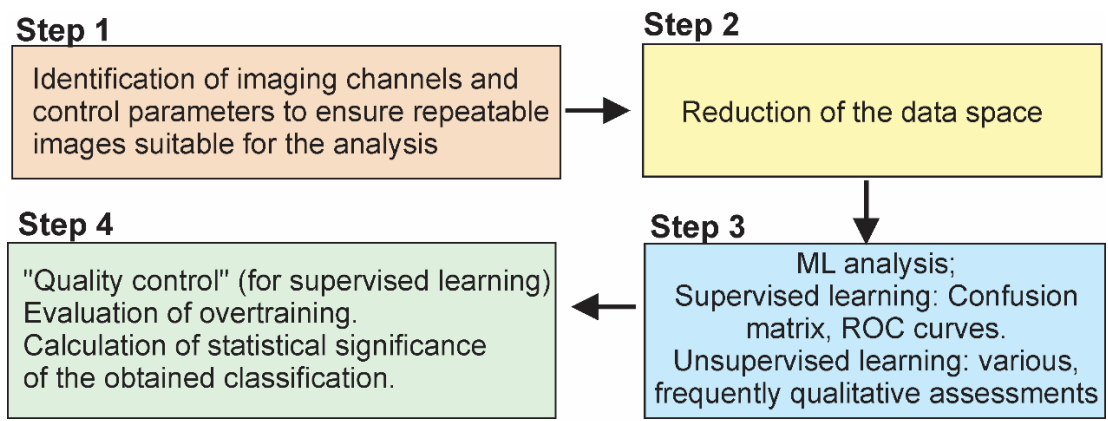

**Figure 2**. A schematic of the recommended steps specific to the ML analysis of AFM images.

### *Example of supervised ML classification of AFM cell images*.

Here, we describe an example of the identification of a specific cell type mainly based on the results reported in ref. [13]. Two similar but genetically altered human colon cancer cell lines, HT29 and Csk cell lines, were studied. Csk cells were a more aggressive cancer line compared to H7-29. Figure 3 shows a representative example of optical and AFM images of these cells. Five different AFM channels are shown, demonstrating cell topography and the maps of four different physical properties, which were recorded in PeakForce QNM and Ringing (RM) modes. The implementation of the steps shown in Figure 2 for this specific example is presented in Figure 4. There are two classes of cells to identify: HT29 and Csk.

Step 1: Each cell was imaged in five different channels simultaneously: the height of the cell (height channel), the adhesion between the AFM probe and cell surface (adhesion channel), and three Ringing mode channels: RM adhesion (RM1 channel), restored adhesion (RM2 channel), viscoelastic losses (RM3 channel). These channels were identified as the least dependent on the imaging parameters and provided absolute values of the image parameters. (Note that the current version of Ringing Mode has 8–9 channels suitable for this purpose). In particular, it was shown that combining four channels in the ML algorithm allowed an increase in the accuracy of classification compared to each individual channel (94% vs. 79, 89, 90. 89%) and a decrease in the dispersion of the distribution of the area under ROC curve (or AUC, the most assumption-independent parameter of classification) substantially.

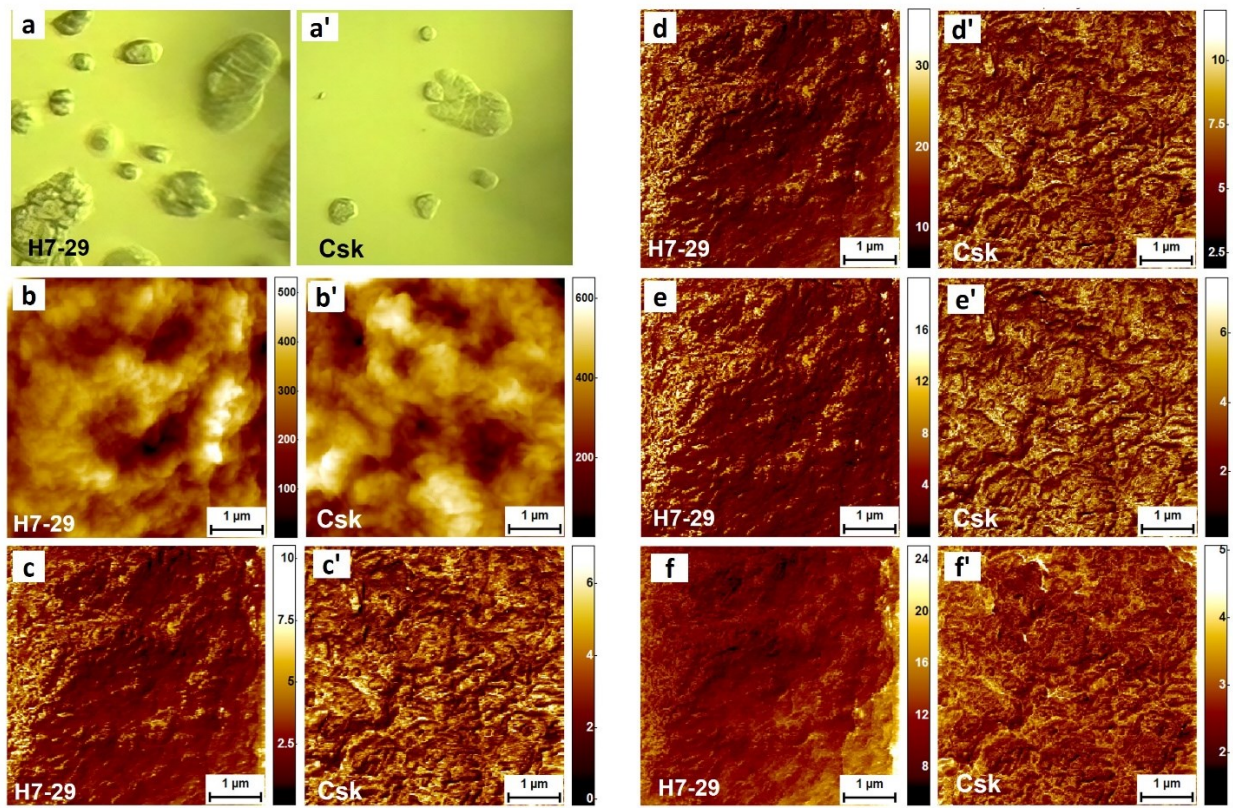

**Figure 3**. Representative images of cells of two classes: highly (H7-29) and less (Csk) aggressive colorectal cancer cell lines. (a,a') 350 x 350 μm$^2$ optical (bright-field) images of cells. (b–f, b'–f ' ) AFM images of both types of cells: (b,b') height, (c,c') adhesion, (d,d') RM adhesion, (e,e') RM restored adhesion, and (f,f') RM viscoelastic adhesion of H7-29 and Csk cell lines, respectively. [Taken from Ref. [13].]

Step 2: Following [11, 13, 50], the images were converted into a set of surface parameters used in engineering to characterize various surfaces. Each parameter is an algorithm or function that converts the image into a single number. A well-known example of such parameters is the surface roughness. The other parameters include various characteristics of directionality, fractal properties, estimation of the size of typical features, etc. This substantially reduces the information for each image. Instead of 512x512 pixels, each image can now be characterized with typically 35-40 surface parameters. Further reduction was performed using two additional algorithms [11, 13]. The first algorithm is based on the analysis of crosscorrelation between the surface parameters. Strongly correlated parameters can be excluded to eliminate the double counting of similar parameters. The second algorithm is the evaluation of the contribution of each parameter to the classification. It can be done, for example, by using the Gini-importance index method, which allows ranking the parameters by their importance for classification. As a result, one can keep only the parameters with the highest importance index. For example, using these two algorithms, only ten parameters were used to describe each cell image.

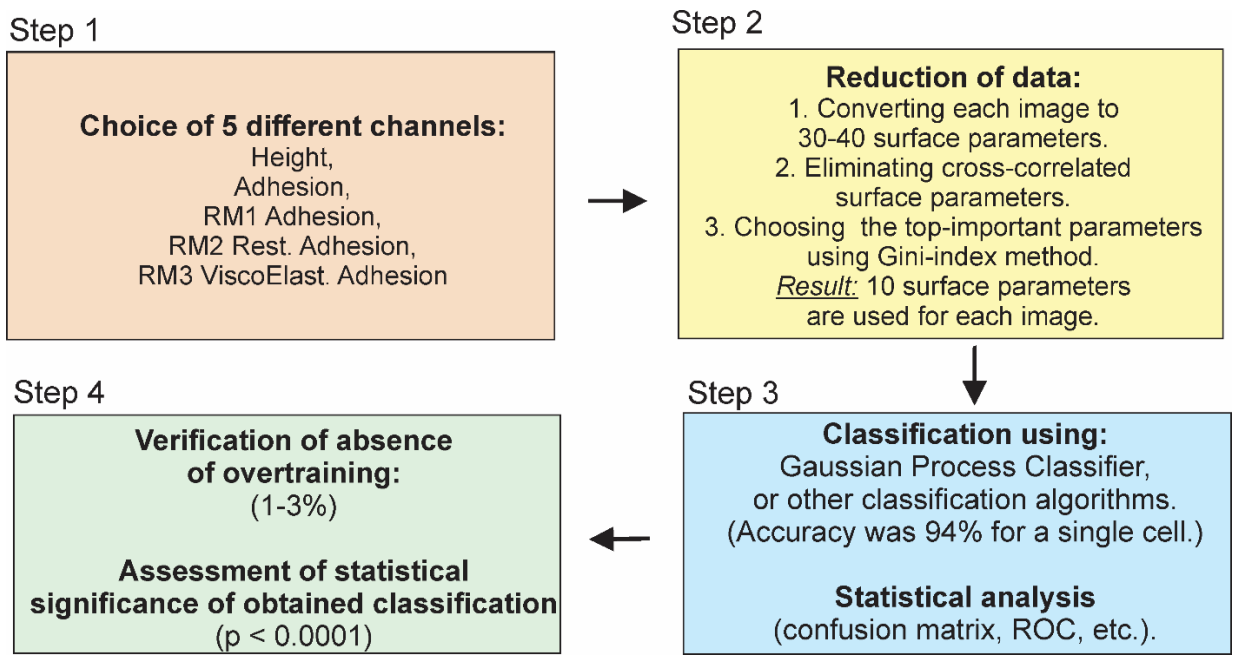

**Figure 4**. An example of a supervised ML method applied to the classification of cancerous cells. It is a specific implementation of the steps presented in Figure 2.

Step 3: The data is randomly split into training and testing subsets. The training of the chosen ML algorithm is done on the training subset, whereas all accuracies (components of the confusion matrix) and ROC curves are calculated on the testing subset. Some algorithms, for example, based on neural networks, may require fine-tuning of the algorithms and repeat the procedure to obtain better accuracy on the testing subsets. Most of the decision trees and regression algorithms do not require this step. It allows to substantially decrease possible overtraining of the developed algorithms.

To avoid the dependence on a particular choice of the training and testing subsets, the split and training are repeated multiple times (can be several hundred; this is called K-fold validation). This allows the calculation of not only the average statistics (area under the ROC curve and the confusion matrix) but also its possible variations (it will be used to find the statistical significance of the obtained classification;

see Step 4). As to a particular ML algorithm, the Gaussian process classifier was used in this example. A particular choice of this algorithm was dictated by the structure of the data (clustering of data submitted by a particular manifold). However, the use of other algorithms gave essentially the same result (unpublished). As an example, the accuracy of the definition of cell type was 94% at the single-cell level when combining four AFM channels.

Step 4: We describe this step in more detail because it is important, and nevertheless, not broadly used in the literature. Overtraining is one of the major bottlenecks of ML algorithms. It leads to an overestimation of the accuracy. It can be found via a substantial decrease in accuracy if one tries to apply the obtained method to a new set of data. As we discussed, it is not easy to generate a lot of AFM images. So, instead of playing with an extended data set, we previously suggested another method to estimate the overtraining [11, 13]. We suggested randomizing the class assignment of the analyzed images by, say, assigning class A or B to each image with an equal probability of 50%. If the used ML algorithm does not have any overtraining problem, the accuracy of identification of a particular class should also be 50% (or the area under the ROC curve is equal to 0.5). If the algorithm demonstrates a higher accuracy than 50%, it is the manifestation of overtraining. And its deviation from 50% is the measure of overtraining. In the examples described in papers, [11, 13] the amount of overtraining was at the level of 1-3 % (this is significantly less than the found accuracy of the data with the actual class assignment, 80-94%).

To find the statistical significance of the obtained classification, we suggested comparing the obtained classification results to the one obtained with the randomized classification described above. It makes sense to use the area under the ROC curve (AUC) because it is probably the most robust description of the power of classification. Nevertheless, the simile approach can be done for any characteristics derived from the confusion matrix. If one uses the data of ref. [13], one has a classical formulation of the statistical analysis for two distributions of AUC. Figure 5 shows these two distributions for the case of four combined channels (see, ref. [13] for detail). One distribution shows the classification of the actual data (the sought signal is present), while the other one presents the distribution of randomized class assignments (the sought signal is not present). Now, one can assess the statistical significance of the difference in the mean values of these two distributions. For the particular example discussed here, the statistical significance was at a level of $p < 0.0001$.

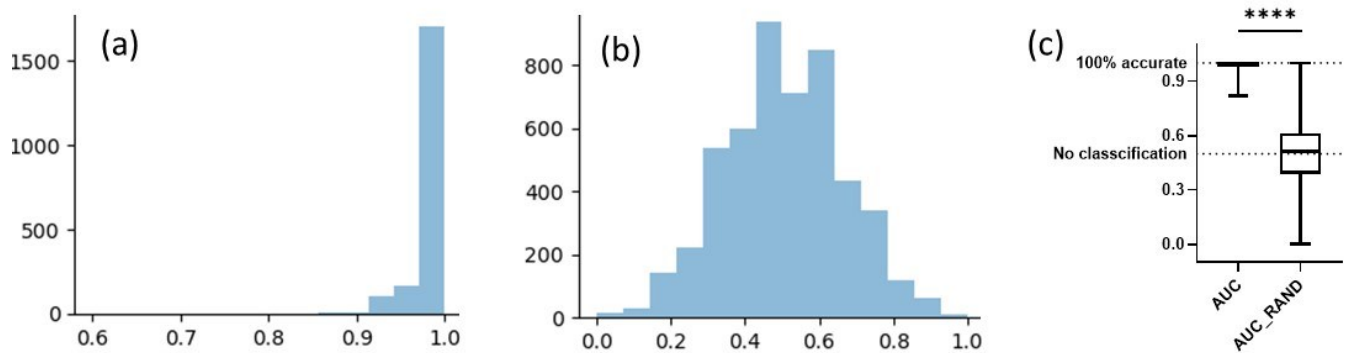

**Figure 5**. Two distributions of the area under the ROC curve (AUC): a) The case of identification of a particular cell type (class A or B) using the actual data of ref. [13]. Multiple AUCs are obtained for the different splitting of the data into training and testing subsets. (b) The same ML algorithms are applied for the case when all AFM images are randomly assigned to either class A or B. Multiple AUCs are obtained here for both different splittings of the data into training and testing subsets and multiple randomizations of the class assignment. (c) shows the statistical presentation of the AUC. The square presents the standard deviation, the middle bar is the mean value of the distribution; the error bars show the limit of the entire distribution (0-100%). **** Corresponds to the statistical significance at the level of p<0.0001.

In conclusion, it is worth adding a few words about unsupervised ML analysis of AFM images. Unsupervised analysis is not typically used for image classification. Unsupervised methods can be extremely helpful for a generic data analysis regardless of the classification. For example, k-means clustering was used in [28] to study the phase transitions and their temperature dependence in ferroelectric materials using AFM images. Unsupervised methods could also be used to reduce the dimension of the data space. As mentioned previously, unsupervised learning analysis is typically highly specific to a particular sample. That is why the main focus of this Perspective is on the supervised ML methods.

The second point to emphasize is that this Perspective avoided the description of the use of CNN, though it is a quite general method currently used to classify images. This is done on purpose because 1) a review of CNN methods applied to AFM has been recently published [16], and 2) CNN and other deep learning methods require large databases for training (see the "rule of ten" mentioned in the text). AFM has difficulty generating the required large databases. The method described here is based on the use of non-deep learning methods, which are typically easier and faster to train. A large variety of available machine learning algorithms makes this approach versatile. It is useful to note that there is a powerful feature of CNN methods, its ability to generate "heatmaps", highlighted geometrical locations of the surface features mostly responsible for the classification. Nevertheless, the generation of the heatmaps was recently introduced for the non-CNN methods as well [38].

A separate question is about the limits of applicability of the described method. Although the approach by itself is quite versatile, there may be some limitations on the type of images that is capable of classifying. We can speculate that the classification of easily recognized objects, like images of cats and dogs, is more effective with CNN. Separation of images that do not have clearly identified large features (like fractal-like images of cells, various textures) would probably be better suited with the approach described here. This topic will be investigated in the future.

**Acknowledgment**: support of this work by NIH R01 CA262147, MLSC Bit-to-Bites program, and NSF CMMI 2224708 grants is acknowledged.